\documentclass[conference]{src/IEEEtran}
\IEEEoverridecommandlockouts
\usepackage{cite}
\usepackage{amsmath,amssymb,amsfonts}
\usepackage{algorithmic}
\usepackage{graphicx}
\usepackage{textcomp}
\usepackage{xcolor}
\usepackage{epsfig}
\usepackage{subfigure}
\usepackage{booktabs}
\usepackage{float}
\usepackage{caption} 
\usepackage[utf8]{inputenc}

\def\BibTeX{{\rm B\kern-.05em{\sc i\kern-.025em b}\kern-.08em
    T\kern-.1667em\lower.7ex\hbox{E}\kern-.125emX}}

\begin{document}
\title{Semi-Automatic Crowdsourcing Tool for Online Food Image Collection and Annotation}
\author{\IEEEauthorblockN{Zeman Shao, Runyu Mao and Fengqing Zhu}
\IEEEauthorblockA{\textit{School of Electrical and Computer Engineering} \\
\textit{ Purdue University}\\
West Lafayette, Indiana, USA \\
\{shao112, mao111, zhu0\}@purdue.edu}
}

\maketitle

\begin{abstract}
Assessing dietary intake accurately remains an open and challenging research problem. In recent years, image-based approaches have been developed to automatically estimate food intake by capturing eat occasions with mobile devices and wearable cameras. To build a reliable machine-learning models that can automatically map pixels to calories, successful image-based systems need large collections of food images with high quality groundtruth labels to improve the learned models. In this paper, we introduce a semi-automatic system for online food image collection and annotation. Our system consists of a web crawler, an automatic food detection method and a web-based crowdsoucing tool. The web crawler is used to download large sets of online food images based on the given food labels. Since not all retrieved images contain foods, we introduce an automatic food detection method to remove irrelevant images. We designed a web-based crowdsourcing tool to assist the crowd or human annotators to locate and label all the foods in the images. The proposed semi-automatic online food image collection system can be used to build large food image datasets with groundtruth labels efficiently from scratch.

\end{abstract}


\section{Introduction}
Many chronic diseases, including cancer, diabetes and heart disease, are closely associated with dietary intake~\cite{Liese_2015,reedy2014higher}. Obtaining accurate, quantitative daily consumption of energy and nutrient is an open research problem.
Although food images with a range of portion sizes have been incorporated into web or mobile applications for dietary assessment methods such as FFQ~\cite{forster2014online, wong2008evaluation, kristal2014evaluation}, dietary record~\cite{wilken2013children}, and 24-h dietary recall~\cite{forster2014online, kirkpatrick2014performance, foster2008children}, these food images merely serve as a visual guidance, and may or may not represent an exact replication of foods consumed by the user. Recent image-based approaches integrating  mobile and wearable technologies have been developed to address the challenge of automated dietary monitoring, such as the Technology Assisted Dietary Assessment (TADA\texttrademark) system~\cite{D-bib:zhu2010, zhu-2015}, FoodLog~\cite{foodlogA}, FoodCam~\cite{joutou2009}, DietCam~\cite{kong2012} and Im2Calories~\cite{Meyers_2015_ICCV}. 

Advances in machine learning, particularly deep learning~\cite{dlnature} techniques such as Convolutional Neural Networks (CNN)~\cite{lecun_cnn}, have shown great successes in many computer vision tasks such as image classification~\cite{krizhevsky2012imagenet, vgg,resnet}, object detection ~\cite{ren2015faster, redmon2017yolo9000} and image segmentation~\cite{mask_rcnn}. 
The success of deep learning methods depend largely on the quantity and quality of data. In general, increasing the size of training data improves the performance of the system. Thus a large dataset with high quality groundtruth labels is always preferred. The groundtruth labels, derived from observable data, is the objective verification of particular properties of an image, used to test the accuracy of computer vision tasks. 

Deep learning based approaches have also been widely used to analyze food images to assess dietary intake in recent years~\cite{Meyers_2015_ICCV, yanai_2017,icip2018, fang2019end}.
Several publicly available food image datasets have been used to validate these approaches.
Although these datasets contain large number of food images sourced from the Internet, several aspects of these datasets can be improved.
In~\cite{chen2009} and~\cite{beijbom2015menu}, the food labels are limited to either popular fast foods or from specific restaurant menus.
In~\cite{cioccaJBHI}, food images are captured in controlled laboratory environment using the same canteen tray and plate. 
In~\cite{bossard14} and~\cite{wang2015recipe}, many images contain incorrect food labels. 
In~\cite{farinella2014benchmark} and~\cite{pouladzadeh2015foodd}, the datasets are designed for food recognition purpose, no food location (i.e., pixels corresponding to foods) information is provided.
In~\cite{kawano2014automatic}, only food item associated with single food label in each image is labeled even if there are food items associated with multiple food labels in the same image.

Collecting food images with proper annotations in a systematic way is a challenging problem. Crowd-sourcing platform, such as Amazon Mechanical Turk (AMT)~\cite{amtwebsite}, is commonly used to select relevant images retrieved from the Internet, and provide simple annotation such as bounding boxes to indicate the pixel location of the food objects. 
For example, in~\cite{kawano2014automatic}, AMT is used as the crowd-sourcing service to select relevant images and add bounding boxes to the selected food images.
In~\cite{rabbi_15}, a machine learning method is proposed to identify high performing worker on AMT in order to achieve high accuracy.
However, the use of AMT is time consuming and expensive. Depending on the type of annotation task,  it can be quite tedious for the worker to accomplish the task in a timely manner while maintaining the quality of the annotation. Therefore, it is desirable to develop an efficient process that can at least automate part of the task and alleviate some of the burden placed on human annotators. 

In this paper, we developed a semi-automatic crowdsourcing system to collect and annotate large sets of online food images. Our system is capable of automatically retrieving relevant images for a particular food label with high accuracy. We also designed a web-based crowdsourcing tool to provide fine-grain annotation of the food images including food labels and food localizations.

\section{Semi-Automatic Food Image Collection System}
 \subsection{Web Crawler}
Online sharing of food images is quickly gaining popularity in recent years on websites such as Facebook, Flickr, Instagram for social networking and Yelp, Pinterest for product review and recommendation. There are also websites dedicated to the sharing of food images, such as yummly and foodgawker. There are hundred-thousands of food images uploaded by smartphone users to these websites. 
Online food images also provide valuable contextual information which is not directly produced by the visual appearance of food in the image, such as the users' dietary patterns and food combinations~\cite{wang2018context}.
These food images can be either retrieved through Application Programming Interface (API) provided by the website, or searched through Google Custom Search Engine (CSE) API to download the food images based on the search terms.
However, since the number of images retrieved is limited by the API provider, it is difficult to collect a large set of food images for each search term.
To quickly collect large number of online food images, we implemented a web crawler to automatically search on the Google Image website based on selected food labels and download the retrieved images according to the relevant ranking on the Google Image.

\subsection{Automatic Food Detection}
Although there are many online images retrieved by the web crawler based on the food labels,  some of these images are considered as noisy image as they do not contain foods.
These images need to be removed from the dataset before they can be used as raining data. However, it can be expensive and time consuming if relying solely on human annotators.
Recently, region proposal methods and region-based convolutional neural networks have shown great success in object detection task\cite{girshick2014}.
In particular, Faster R-CNN \cite{faster_rcnn}, which uses a Region Proposal Network that simultaneously predicts object bounds and objectness scores at each position, is a popular method for object detect.
We trained a  Faster R-CNN for food region detection to remove non-food images.
The objectness score associated with detected regions is defined as the ``foodness" score.
Non-food images can be removed based on the "foodness" score detected in an image. 
A threshold is needed to decide whether to keep or discard the retrieved food image.
To best determine this threshold value, we used a statistical indicator to measure the performance of different threshold values for a subset of the food images retrieved from the Internet. Details are discussed in Section~\ref{section:auto}.

\subsection{Crowdsourcing Tool}
We have previously designed and developed a crowdsourcing tool for online food image identification and segmentation, and showed its efficiency and effectiveness in locating food items and creating groundtruth segmentation masks associated with all the foods presented in an image in~\cite{fang_2018_ssiai}.
We integrated the food item localization functionality into our semi-automatic data collection system, and made additional improvements in order to further reduce the annotation time requirement for the crowd workers.
The components of the crowdsouring tool and processes of using the tool are described in details below.

\subsubsection{Noisy Image Removal}
\begin{figure}[h]
\centering
\vspace{-0.15 in}
\includegraphics[scale=0.5]{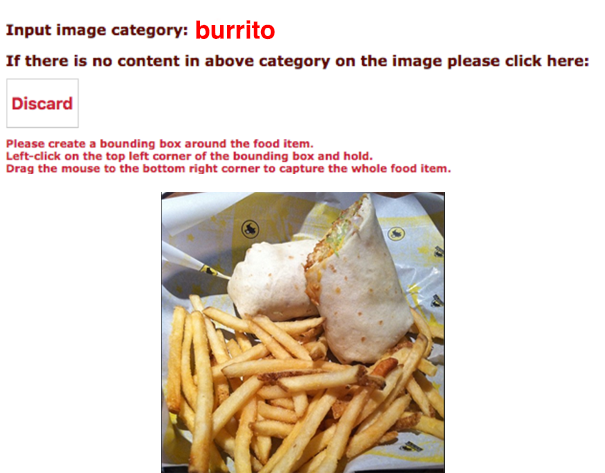}
\caption{An example of confirming the food label associated with an image.}
\vspace{-0.15 in}
\label{fig:localization}

\end{figure}
\begin{figure}[h]
\centering
\vspace{-0.15 in}
\includegraphics[scale=0.5]{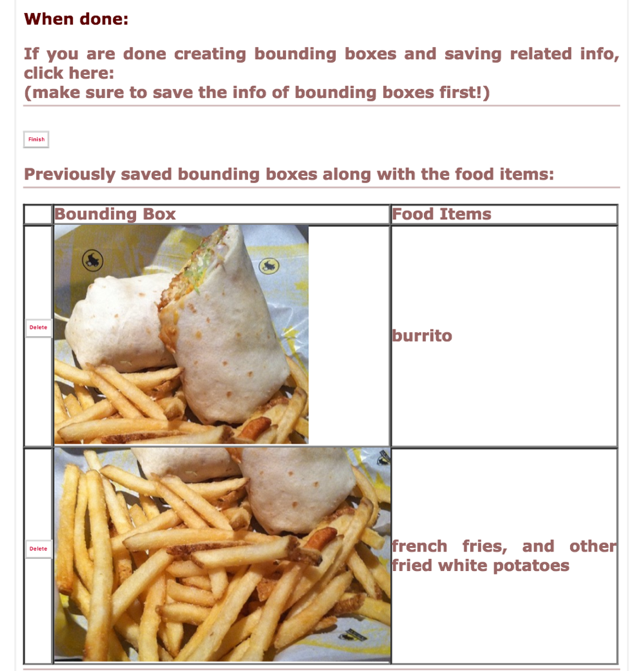}
\caption{An example of reviewing existing bounding boxes for an image.}
\vspace{-0.15 in}
\label{fig:review}
\end{figure}
The food images obtained after automated non-food image removal will inevitably contain noisy image that we cannot use. 
The noisy images are defined as those that either contain irrelevant content which means no relevant food item are present in the image, or have aesthetic appearance.
Food images with aesthetic appearances are likely captured and/or retouched by professional photographers and have different visual appearance compared to images taken by smartphone users in terms of textures, colors, angles and layouts.
A tutorial is provided to the crowds at the beginning of and is available during the annotation task so the crowds can successfully remove images with a set of criteria.
In the tutorial, there are side-by-side comparisons of images and a description of the criteria for noisy image removal. 
We also provided a food label which is used in web crawler to download the image, a one-click confirmation button and a short-cut key on the keyboard to simplify and speed up the process.
An example is shown in Figure~\ref{fig:localization}. 
Based on a preliminary experiment, it takes on average one second for the crowd to examine and remove one noisy image.

\subsubsection{Food Item Localization}
Only food images passed both automatic and crowd-sourced noisy image removal process are assigned for further food item localization.
To locate food items in an image, one needs to draw a bounding box around each food item.
This task can be performed efficiently by click-and-drag using a computer mouse on the web interface. 
A bounding box along with the reference food label is generated, the crowd could change the associated label to any other labels from provided food list.
Once a bounding box is saved, pixels within the bounding box is cropped out of the original image and along with the associated food label is listed below for review.
The crowd can verify or delete an existing bounding box and draw a new one before moving on to the next image, as shown in Figure~\ref{fig:review}.

\section{Experimental Results}
\subsection{Automatic Food Image Detection}\label{section:auto}
We manually verified 1,000 food images from 50 food categories and another 1,000 non-food images as training dataset. 
We noticed that the ratio of food images vs. noisy images vary for different food labels in our dataset we collected.
Therefore, a good threshold value should yield good performances on different ratios of food and non-food mixtures.
We tested image mixtures which have 50-90\% food images with our trained Faster R-CNN network. 
For each mixture, we prepared 1,000 different trials based on stratified sampling so that the results can be more general. 

\begin{figure}[h]
\centering
\vspace{-0.15 in}
\includegraphics[scale=0.35]{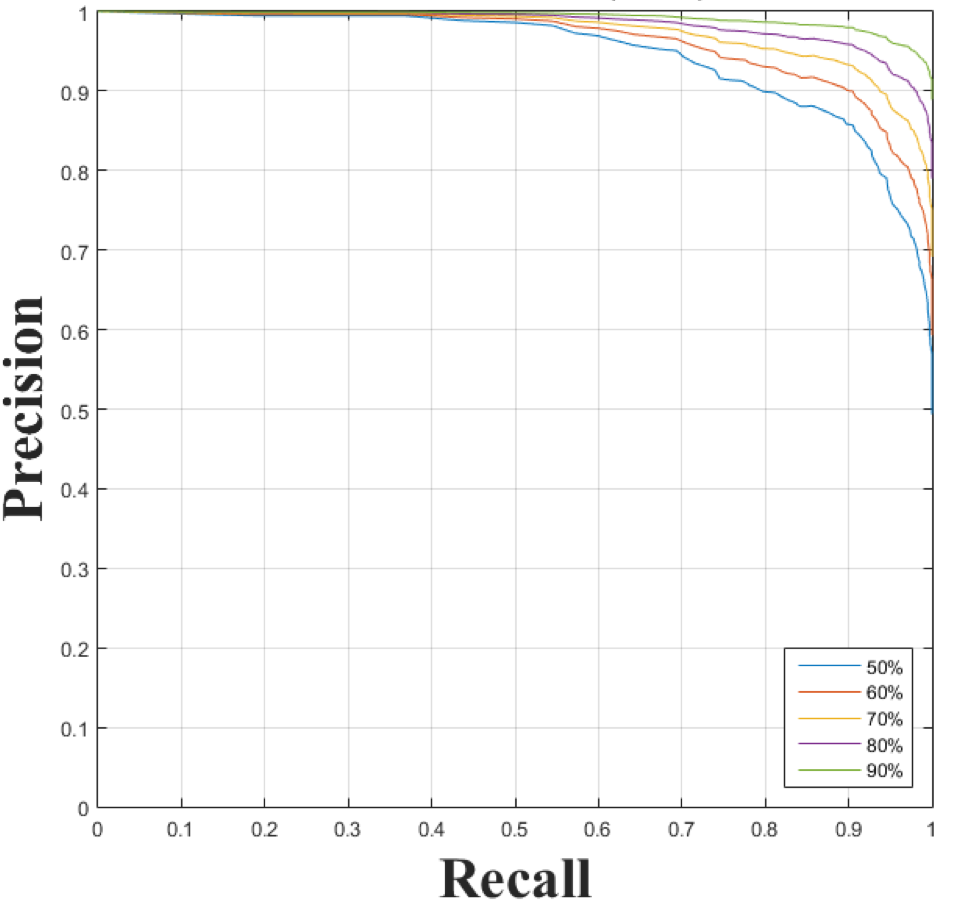}
\caption{PR Curve for Different Food Image Mixtures Ratio}
\vspace{-0.15 in}
\label{fig:prc}
\end{figure}

The Precision-Recall (PR) curves summarize the trade-off between precision (true positive rate) and recall (positive predictive value). We would like to have a curve close to the upper right corner.
As shown in Figure~\ref{fig:prc}, the Faster R-CNN based food image detection has a good performance for different food image mixtures ratio. 
To obtain threshold value that yields high precision and high recall, we record both precision and recall of each ``foodness” score for the different image mixtures ratio.
Since we want to build a large food image dataset, the automatic process may falsely discard no more than 20\% correct images (recall $\geq$ 0.8).
To improve the quality of subsequent crowdsourcing tasks, the remaining images should contain more than 80\% correct images (precision $\geq$ 0.8).
We record the acceptable "foodness" score range that satisfies both precision and recall larger than 0.8 for different food image mixtures ratio in Table~\ref{table:threshold}.
It shows that the range [0.57, 0.77] satisfies our requirements in all 5 different mixtures.

\begin{table}[h]
\centering
\vspace{-0.05 in}
\caption{Acceptable "Foodness" Score Range for Different Food Image Mixtures Ratio}

\begin{tabular}{ |c|c| } 
\hline
Food Image Count & Acceptable "Foodness" Score \\
\hline
50\%&               [0.57, 0.77]\\
60\%&               [0.45, 0.77]\\
70\%&               [0.32, 0.77]\\
80\%&               [0.05, 0.77]\\
90\%&               [0.00, 0.77]\\
\hline
\end{tabular}
\vspace{-0.25 in}
\label{table:threshold}
\end{table}

\subsection{Food Item Localization}
We selected 10 food labels and used our web crawler to download corresponding food images from online sources. We then applied the automatic food detection on the downloaded images to remove noisy image.
In total, there were 3,838 food images uploaded into our online crowdsourcing tool, which were assigned to the crowd to remove noisy images and draw bounding boxes around each food in these images.
After noisy images removal and food item location using the crowdsoucing tool, we obtained 3,058 bounding boxes. Table~\ref{table:boundingbox} shows the number of bounding boxes for each food label.
\begin{table}[H]
\centering
\vspace{-0.05 in}
\caption{Food Image Count and Bounding Box Count for Different Food Label}
\begin{tabular}{ |c|c|c| }
\hline
Food Label              & Food Image Count          & Bounding Box Count    \\
\hline
Doughnut                &1279                       &937 \\      
Cupcake                 &596                        &542 \\
Cornbread               &385                        &269 \\
Tostada                 &364                        &247 \\
Broccoli                &350                        &304 \\
Cookie                  &214                        &160 \\
Waffle                  &200                        &170 \\
Red Wine                &174                        &132 \\
Bananas                 &156                        &157 \\
Cheese Burger           &120                        &140 \\
\hline
\end{tabular}
\vspace{-0.1 in}
\label{table:boundingbox}
\end{table}

\section{Conclusion}
We have designed and implemented a semi-automatic system to collect and annotation online food images based on the food category search terms.
We showed that our system can improve the efficiency and quality of building a large food image dataset with ground truth information such as food category labels and food object locations for each food item in the image.
We are currently using this system to build a large food image dataset which can be used to develop new image-based dietary assessment methods.


\bibliographystyle{IEEEtran}
\bibliography{src/bfndma2019}

\end{document}